\documentclass[prb,twocolumn,longbibliography,amsmath,floatfix,superscriptaddress,amssymb,nobibnotes]{revtex4-2}
\usepackage{tikz}
\usepackage{makeidx,amssymb,amsmath,amsthm,graphicx}
\usepackage[toc,page]{appendix}
\usepackage{dcolumn}
\usepackage{bm}
\usepackage{color}
\usepackage{placeins}
\usepackage{epstopdf}
\usepackage{multirow}
\usepackage{rcs}
\usepackage[normalem]{ulem}
\usepackage{chngcntr}
\usepackage[thinspace,thinqspace]{SIunits}
\usepackage{natbib}
\usepackage{hyperref}

\usepackage[T1]{fontenc}            
\usepackage[latin1]{inputenc}
\usepackage{color}

\usepackage{babel}
\makeatother
\begin{document}

\title{Influence of the Pd-Si ratio on the valence transition in EuPd$_2$Si$_2$ single crystals}

\author{Kristin Kliemt}
\affiliation{%
 Kristall- und Materiallabor, Physikalisches Institut, 
 Goethe-Universit\"at Frankfurt, Max-von-Laue Strasse 1, 
 60438 Frankfurt am Main, Germany
}
\email{kliemt@physik.uni-frankfurt.de}
\author{Marius Peters}
\affiliation{%
 Kristall- und Materiallabor, Physikalisches Institut, 
 Goethe-Universit\"at Frankfurt, Max-von-Laue Strasse 1, 
 60438 Frankfurt am Main, Germany
}%
\author{Isabel Reiser}
\affiliation{%
 Kristall- und Materiallabor, Physikalisches Institut, 
 Goethe-Universit\"at Frankfurt, Max-von-Laue Strasse 1, 
 60438 Frankfurt am Main, Germany
}%
\author{Michelle Ocker}
\affiliation{%
 Kristall- und Materiallabor, Physikalisches Institut, 
 Goethe-Universit\"at Frankfurt, Max-von-Laue Strasse 1, 
 60438 Frankfurt am Main, Germany
}%
\author{Franziska Walther}
\affiliation{%
 Kristall- und Materiallabor, Physikalisches Institut, 
 Goethe-Universit\"at Frankfurt, Max-von-Laue Strasse 1, 
 60438 Frankfurt am Main, Germany
}%
\author{Doan-My Tran}
\affiliation{%
 Kristall- und Materiallabor, Physikalisches Institut, 
 Goethe-Universit\"at Frankfurt, Max-von-Laue Strasse 1, 
 60438 Frankfurt am Main, Germany
}%
\author{Eunhyung Cho}
\affiliation{%
 Kristall- und Materiallabor, Physikalisches Institut, 
 Goethe-Universit\"at Frankfurt, Max-von-Laue Strasse 1, 
 60438 Frankfurt am Main, Germany
}%
\author{Michael Merz}
\affiliation{%
 Institute for Quantum Materials and Technologies (IQMT), Karlsruhe Institute of Technology (KIT), 76344 Eggenstein-Leopoldshafen, Germany
}%
\altaffiliation{%
 Karlsruhe Nano Micro Facility (KNMFi), Karlsruhe Institute of Technology (KIT), 76344 Eggenstein-Leopoldshafen, Germany
}%
\author{Amir Abbas Haghighirad}
\affiliation{%
 Institute for Quantum Materials and Technologies (IQMT), Karlsruhe Institute of Technology (KIT), 76344 Eggenstein-Leopoldshafen, Germany
}%
\author{Dominik C.~Hezel}
\affiliation{%
Institut f\"ur Geowissenschaften, Petrologie und Geochemie, Goethe-Universit\"at Frankfurt, Altenh\"ofer Allee 1, 
60438 Frankfurt am Main, Germany
}%
\author{Franz Ritter}
\affiliation{%
 Kristall- und Materiallabor, Physikalisches Institut, 
 Goethe-Universit\"at Frankfurt, Max-von-Laue Strasse 1, 
 60438 Frankfurt am Main, Germany
}%
\author{Cornelius Krellner}
\affiliation{%
 Kristall- und Materiallabor, Physikalisches Institut, 
 Goethe-Universit\"at Frankfurt, Max-von-Laue Strasse 1, 
 60438 Frankfurt am Main, Germany
}%


\date{June 1, 2022}

\begin{abstract}
Single crystals of intermediate valent EuPd$_2$Si$_2$ were grown from an Eu-rich melt by the Bridgman as well as the Czochralski technique. The chemical and structural characterization of an extracted single crystalline Czochralski-grown specimen yielded a slight variation of the Si-Pd ratio along the growth direction and confirms the existence of a finite Eu(Pd$_{1-m}$Si$_m$)$_2$Si$_2$ homogeneity range. The thorough physical characterization carried out on the same crystal showed that this tiny variation in the composition affects the temperature $T_v$ at which the valence transition occurs. These experiments demonstrate a strong coupling between structural and physical properties in the prototypical valence-fluctuating system EuPd$_2$Si$_2$ and explain the different reported values of $T_v$. 

\end{abstract}

\maketitle


\section{\label{sec:level1}Introduction}

The study of emergent phenomena due to strong electron-electron interactions is at the heart of research in condensed matter. For these type of systems, the coupling of electronic degrees of freedom to the lattice are usually not that much in the focus of research. In recent years, however, more and more physical  phenomena have arisen in materials which additionally show a strong coupling of the electronic degrees of freedom to the lattice. 
Examples are tetragonal iron-based superconductors, where a so-called collapsed tetragonal phase appears upon cooling or application of pressure  
which is connected to an abrupt change in the $c$-lattice parameter \cite{Kreyssig2008}. In the organic conductor
 $\kappa$-(BEDT-TTF)$_2$Cu$[$N(CN)$_2]$Cl, the breakdown of Hooke's law was observed 
in a certain pressure-temperature range at the Mott-metal insulator transition and the appearance of 
critical elasticity was proposed \cite{Gati2016}.
In the past, intermetallic Eu systems were studied due to the intermediate valent nature 
of Eu in these compounds. The change of the Eu valence can be tuned by 
temperature \cite{Mimura2004} and/or pressure \cite{Adams1991}.
Nowadays, these compounds attract attention due to 
the large volume changes accompanied by the valence transition. 
Eu can occur in different valence states which are very close in energy. 
The magnetic Eu$^{(2+\delta)+}$ ($4f^7$ configuration) leading 
to a larger unit cell volume, appears at low pressures and high temperatures whereas the non-magnetic 
Eu$^{(3-\delta^{\prime})+}$ ($4f^6$ configuration) leading to a smaller unit cell volume
 is adopted at high pressures and low temperatures.
In a general phase diagram, the two regions are separated at low temperatures by a line
of first-order valence transitions \cite{Onuki2017}. Of special interests are intermediate-valent Eu-122 compounds which
can be tuned by pressure (temperature) towards the second order critical endpoint 
of the line of the first order valence transitions.
In particular, EuPd$_2$Si$_2$ is proposed to be located very close to the 
critical pressure in the generic phase diagram. 
The extrapolation of results obtained on polycrystalline samples yielded that 
EuPd$_2$Si$_2$ is located on the high-pressure side of the critical endpoint 
\cite{Batlogg1982, Schmiester1982}.
It is expected that in EuPd$_2$Si$_2$ critical elasticity meaning 
a strong lattice softening and non-linear strain-stress relations  might occur as well.

First reports on EuPd$_2$Si$_2$ appeared in the 1980's where the intermediate-valent nature of Eu was studied.
Here, a strong temperature dependence of the $^{151}\rm Eu$ M\"ossbauer isomer shift together 
with a maximum in the temperature dependence of the magnetic susceptibility was observed for the first time 
in a Eu-based intermediate valent compound \cite{Sampathkumaran1981}. 
The rare earth ion undergoes a valence transition from Eu$^{2.2+}$ at room temperature to Eu$^{2.9+}$ in the bulk 
at low temperature $\leq 100\,\rm K$ while photoemission spectroscopy revealed 
divalent Eu at the surface in the whole temperature range \cite{Sampathkumaran1981, Martensson1982, Mimura2004}.
The increase of the Eu valence in the bulk upon cooling strongly influences the lattice parameters. 
In particular, it causes a change of the $a$ parameter while the $c$ parameters remains nearly unchanged. 
The analysis of the Bragg line-widths shows a broadening of the ($2\, 0\,0$) and therefore suggested 
the existence of a temperature dependent distribution of $a$ parameters in powder samples \cite{Jhans1987}.
A valence transition can be induced not only by decreasing the temperature but also by applying pressure 
which was revealed first by electrical resistivity and thermoelectric power measurements \cite{Vijayakumar1981}. 
Later on, from the temperature as well as the pressure dependence of the electrical resistivity, the 
linear compressibility and the thermal expansion a pressure-temperature (p-T) phase diagram was 
constructed by Batlogg {\it et al.} \cite{Batlogg1982}. From these data it was concluded that 
EuPd$_2$Si$_2$ shows a continuous phase transition and is situated near the critical endpoint 
of a line of first order phase transitions in the p-T diagram. 
It was proposed that this endpoint can be reached by applying low negative chemical pressure 
for instance via substitution of the smaller Pd atoms by larger Au atoms \cite{Croft1982}.
It was shown by M\"ossbauer effect measurements that the transition becomes first order 
when applying negative chemical pressure as for instance for $x>0.1$ in 
Eu(Pd$_{1-x}$Au$_x$)$_2$Si$_2$ \cite{Segre1982}.
EuPd$_2$Si$_2$ was studied by neutron scattering and by measurements of the magnetic susceptibility. 
It turned out that the tail in the static susceptibility at low temperatures can be caused by 
the ordering of a magnetic secondary phase \cite{Holland1987}.
Although EuPd$_2$Si$_2$ is one of the prototypical valence-fluctuating systems, previous studies were mostly carried out on polycrystalline samples. It should be mentioned 
that the crystal growth of Eu compounds is challenging due to the high vapour pressure of Eu 
and its affinity to oxygen. Only recently small single crystals of EuPd$_2$Si$_2$ grown by the 
Bridgman method became available \cite{Onuki2017}. 
The comparison of poly and single crystal heat capacity data from different sources \cite{Wada2001, Onuki2017} shows sample dependencies concerning 
the temperature at which the valence transition occurs. From this fact one might speculate that 
EuPd$_2$Si$_2$ is a compound that exists in a homogeneity range as it was already found for the related material ErPd$_2$Si$_2$ \cite{Mazilu2008}. 
In \cite{Kuzhel2010} an EuPd$_{2-x}$Si$_{2+x}$ alloy series was studied; the respective samples were prepared by arc-melting followed by an annealing step. The samples studied contained either secondary phases or grain boundaries where magnetic impurities accumulated. Due to this fact, the information that can be gained from their magnetic characterization is limited.
This motivates further attempts to grow single crystals with a minimum of magnetic inclusions 
which enable studies of the physical properties in connection with the study of the structural details of the material 
in a reliable manner.

We used the Czochralski method to provide large and pure single crystals of EuPd$_2$Si$_2$ 
and studied the influence of the Pd-Si composition change in a sample on its physical properties.
In future studies, this will help figuring out whether the scenario of critical elasticity \cite{Garst2015},
which was recently proposed for the Mott transition in $\kappa$-(BEDT-TTF)$_2$Cu$[$N(CN)$_2]$Cl,
can be found in EuPd$_2$Si$_2$.

\section{Experimental Details}
\begin{figure}
\centering
\includegraphics[width=0.5\textwidth]{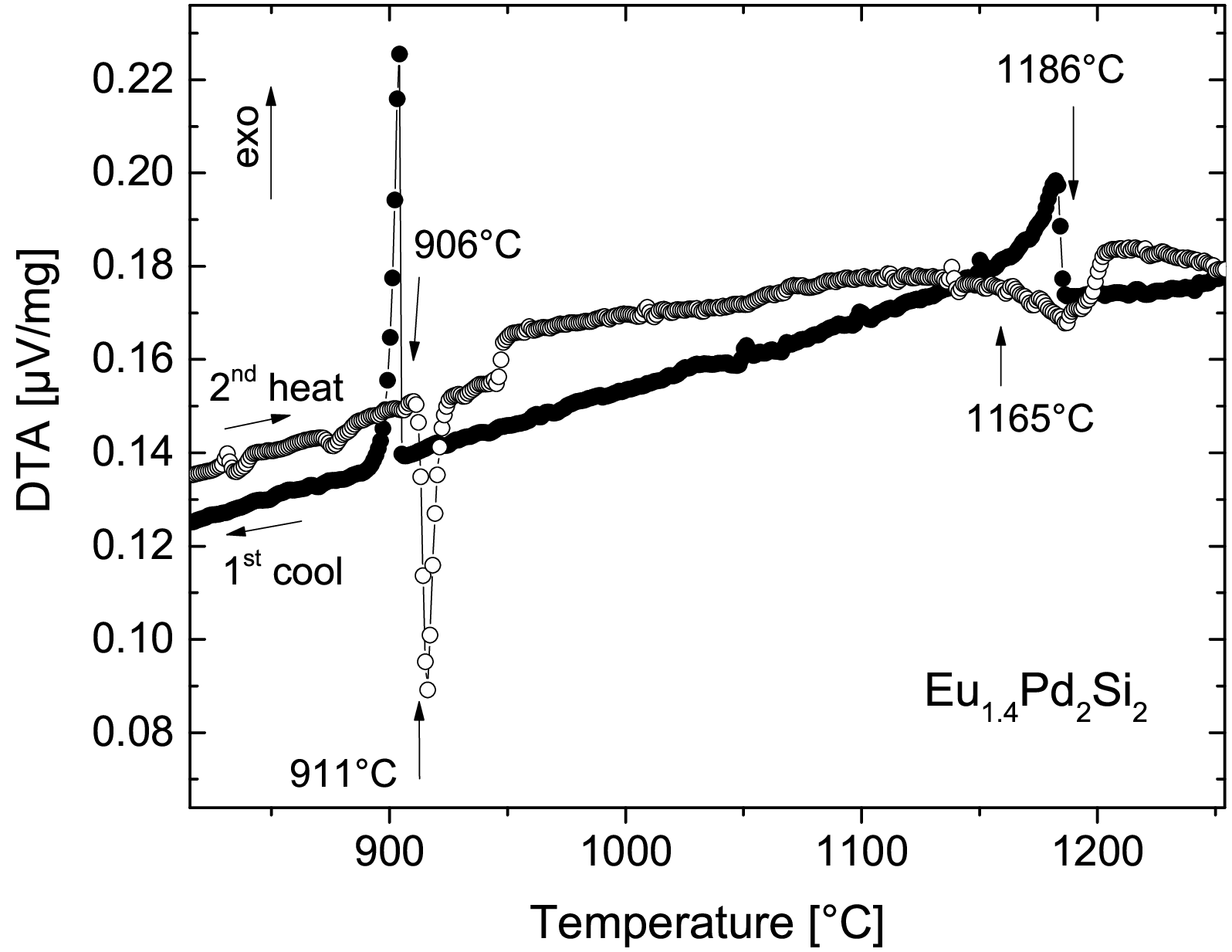}
	\caption[]{DTA experiment performed for the initial stoichiometry of 
Eu$\,:\,$Pd$\,:\,$Si $= 1.4\,:\,2\,:\,2$ (Eu$_{1.4}$). Two signals were observed upon cooling (closed circles)/heating (open circles)  corresponding to the solidification/melting of EuPd$_2$Si$_2$ at around $1180^{\circ}\rm C$ and the remaining flux at lower temperatures.}
\label{DTA_EC05}
\end{figure}
A Simultaneous Thermal Analysis device (STA 449 C, Netzsch), which allows simultaneous thermogravimetry (TG) 
and differential thermal analysis (DTA) was used to get insight in the  solidification process of an Eu-enriched starting stoichiometry.
DTA under Ar flow of $150\,\rm ml/min$ was done with a heating/cooling rate of $10\,\rm K/min$ using the 
different stoichiometries Eu$\,:\,$Pd$\,:\,$Si $= 1.4\,:\,2\,:\,2$ (Eu$_{1.4}$, Fig.~\ref{DTA_EC05}) 
and Eu$\,:\,$Pd$\,:\,$Si $= 1.5\,:\,2\,:\,2$ (Eu$_{1.5}$) in Al$_2$O$_3$ and graphite crucibles. We observed a strong evaporation of Eu 
leading to a weight loss of up to $2.5\%$
and a reaction with the Pt wires of the thermocouple. Furthermore, we found that the Al$_2$O$_3$ 
crucible was attacked by the melt and that Al was incorporated in the sample. 
The DTA signal of the second cooling from a run in a graphite crucible (sample EC05), Fig.~\ref{DTA_EC05}, shows two features. One which indicates the solidification of EuPd$_2$Si$_2$ 
in the flux at 1186$^\circ$C for the Eu$_{1.4}$ sample (1205$^\circ$C for Eu$_{1.5}$) 
and a second which belongs to the solidification of the flux itself at 906$^\circ$C.
For the crystal growth from a Eu-rich melt, high purity elements Eu (99.99\%, chunks, Alpha Aesar), 
Pd (99.999\%, rod, Heraeus), Si (99.99\%, pieces, Cerae) were used in a ratio of 
Eu$\,:\,$Pd$\,:\,$Si $=1.5\,:\,2\,:\,2$ as proposed by \cite{Onuki2017}.
The Bridgman growth was done from the elements in a vertical resistive furnace 
(GERO HTRV70-250/18). Since the crucible material tantalum was attacked by the melt, we used an additional graphite inner crucible.

The Czochralski growth was done using prereacted material with a total weight of up to $15\,\rm g$. 
The initial stoichiometry varied from Eu$_{1.2-1.5}$Pd$_2$Si$_2$.
In the first step, the binary compound PdSi ($\approx 900^\circ$C) was prepared by arc melting, to reduce the high melting points T$_{\rm m}$ of Pd (1555$^\circ$C) and Si (1414$^\circ$C). Afterwards, using a glove box,
this material together with elementary Eu (T$_{\rm m} = 822^\circ$C) was put in a glassy carbon inner crucible 
and enclosed in a niobium outer crucible. We used a box furnace (Linn) 
to heat the precursor up to 835$^\circ$C under Ar flow. 
The Czochralski crystal growth was done in an Arthur D. Little growth chamber equipped with a high-frequency generator from H\"uttinger. 
The chamber was filled with an Ar pressure of 20 bar to slow down the evaporation of Eu at high temperatures. 
The air-sensitive precursor was quickly put in a cold copper crucible of a levitation setup followed by an extended purging to remove oxygen. This setup enables a quasi crucible-free 
crystal growth from a levitating melt which is necessary due to the strong reaction of the melt with any tested crucible material.
For this study, samples from three different Czochralski growth  experiments $\#1$ (MP401), $\#2$ (MP413a) and $\#3$ (MP413b) were investigated.
We used powder X-ray diffraction (PXRD, Cu-K$_{\alpha}$ radiation) to check the lattice parameters of crushed single crystals and their T dependence, and energy- as well as wavelength-dispersive X-ray spectroscopy (EDX, WDX) to determine the chemical composition of the crystals. \\
EDX measurements were performed at the Kristall- und Materiallabor, Goethe Universit\"at Frankfurt, using a scanning electron microscope, Zeiss DSM 940 A, with an additional energy dispersive detector (EDAX Ametek GmbH). At the KIT, the chemical composition of the crystals was examined by energy-dispersive x-ray spectroscopy employing a Coxem EM-30$^N$ benchtop scanning electron microscope equipped with an Oxford Instruments detection system.
WDX measurements were performed with a JEOL-8530F Plus Hyperprobe at the Institut f\"ur Geowissenschaften, Goethe Universit\"at Frankfurt. Individual points were measured with $3\,\mu\rm m$ spot sizes at an acceleration voltage of $20\,\rm kV$ and a current of $30\,\rm nA$, with counting times of $30\,\rm s$ (peak) and $15\,\rm s$ (background). Pure Si and Pd metals as well as a Eu-Phosphate (EuPO$_4$) were used as standards. As some oxidation of the standards cannot be excluded, some minor systematic shift of the measured concentrations (typically below $0.2\,\rm rel.\%$ for Pd and Si, and up to $1.5\,\rm rel.\%$ for Eu) can not be excluded. Internal reproducibility is typically below $0.5\,\rm  rel.\%$.
For a more detailed structural characterization 
X-ray diffraction (XRD) data on representative Eu(Pd$_{1-m}$Si$_m$)$_2$ single-crystal samples were collected at 295 K on a STOE imaging plate diffraction system (IPDS-2T) using Mo $K_{\alpha}$ radiation. All accessible reflections ($\approx 5000$) were measured up to a maximum angle of $2 \Theta =65^\circ$. The data were corrected for Lorentz, polarization, extinction, and absorption effects. Using SHELXL \cite{Sheldrick2008} and JANA2006 \cite{Petricek2014}, all averaged symmetry-independent reflections ($I > 2 \sigma$) have been included for the respective refinements in the tetragonal space group (SG) $I4/mmm$. For all compositions the unit cell and the space group were determined, the atoms were localized in the unit cell utilizing random phases as well as Patterson superposition methods, the structure was completed and solved using difference Fourier analysis, and finally the structure was refined. In all cases the refinements converged quite well and show excellent reliability factors (see GOF, $R_1$, and $wR_2$ in Table \ref{tab:table1}). 

\begin{table}[!ht]
\tabcolsep=0.01cm
	{\small
	\begin{tabular}[t]{llccc}
            &                           &       $E0$          &          $E4$      &       $E8$ \\ \hline
			& SG                        &     $I4/mmm$         &      $I4/mmm$         & $I4/mmm$    \\
			&  $a$ (\AA)              &      4.2392(6)       &        4.2406(5)    &       4.2408(7) \\
			&  $c$ (\AA)              &      9.8674(12)     &       9.8690(11)   &       9.8631(12) \\
			&  $\alpha$ (\textdegree)             &       90             &          90         &      90   \\
			&  $\beta$ (\textdegree)              &       90             &          90         &      90   \\
			&  $\gamma$ (\textdegree)             &       90             &          90         &       90 \\
			&  V (\AA$^3$)            &      177.3           &         177.5       &       177.4 \\ 
    Eu      & Wyck.                     &      $2a$            &         $2a$        &       $2a$ \\
			& $x$                       &        0             &          0          &         0 \\
			& $y$                       &        0             &          0          &         0 \\
			& $z$                       &        0             &          0          &         0 \\
			& $U_{\rm iso}$ (\AA$^2$) &      0.00898(14)     &      0.00786(11)   &     0.00836(12) \\ 
    Pd/Si   & Wyck.                     &       $4d$           &         $4d$        &     $4d$  \\
            & $x$                       &        $\frac{1}{2}$         &        $\frac{1}{2}$        &      $\frac{1}{2}$ \\
			& $y$                       &          0           &           0         &     0 \\
			& $z$                       &        $\frac{1}{4}$ &        $\frac{1}{4}$     &   $\frac{1}{4}$ \\
			& occ. (\%)                 &        97.0/3.0(5)        &        98.1/1.9(4)     &   98.8/1.2(5) \\
			& $U_{\rm iso}$ (\AA$^2$) &  0.00978(13)         &      0.00889(10)    &      0.00943(13) \\
    Si      & Wyck.                     &      $4e$            &         $4e$        &       $4e$ \\
            & $x$                       &        0             &          0          &      0  \\
			& $y$                       &        0             &          0          &      0 \\
			& $z$                       &    0.37783(19)        &      0.37754(14)    &      0.37719(23) \\
			& $U_{\rm iso}$ (\AA$^2$) &    0.01039(35)       &      0.00897(27)       &      0.00947(34) \\
			& GOF                       &       1.90           &         1.47        &       1.38 \\
			& $wR_2$ (\%)               &       4.15           &         3.20        &       3.33  \\
			& $R_1$ (\%)                &       1.72           &         1.24        &       1.21 \\ \hline
            & bond lengths               &                    &                      &           \\
			& Eu-Si (\AA)               &       3.2309(9)      &         3.2330(7)   &       3.2341(11) \\
			& (Pd,Si)-Si (\AA)          &       2.4665(11)      &        2.4657(8)   &       2.4637(13) \\
			& Si-Si (\AA)               &       2.4110(30)      &        2.4170(19)   &      2.4220(28) \\
	\end{tabular}
	}
\caption{\label{tab:table1} {\small
	Crystallographic data of Eu(Pd$_{1-m}$Si$_m$)$_2$Si$_2$ at 295 K for representative samples $E0$, $E4$, and $E8$ determined from single-crystal x-ray diffraction. For all compositions the structure was refined in the tetragonal space group (SG) $I4/mmm$. $U_{\rm iso}$ denotes the isotropic atomic displacement parameters (ADP). The ADPs were refined anisotropically but due to space limitations only the $U_{\rm iso}$ are listed in the table. The Wyckoff positions (Wyck.) are given as well. While for all samples the $4e$ Wyckoff position is completely occupied with Si, a certain amount of up to 3 \% Si is found on the Pd site (Wyckoff position $4d$). Furthermore, some selected bond distances are depicted as well. Errors shown are statistical errors from the refinement.}}
\end{table}	

The samples $E0$, $E4$ and $E8$ for the single crystal analysis were extracted from the Czochralski grown sample $\#3$ at distances of $2.5\,\rm mm$, $7\,\rm mm$ and $11\,\rm mm$ below the seed.
Low-temperature PXRD was done in a Siemens D500 diffractometer (Cu-K$_{\alpha}$ radiation).
A Laue camera with X-ray radiation from a tungsten anode was used to determine the orientation of our 
single crystals and to localize grain boundaries in the Czochralski grown samples. 
Heat capacity, four-point resistivity and magnetization measurements were performed using the 
commercial measurement options of a Quantum Design PPMS.

\section{Results}

\begin{figure}
\centering
\includegraphics[width=0.5\textwidth]{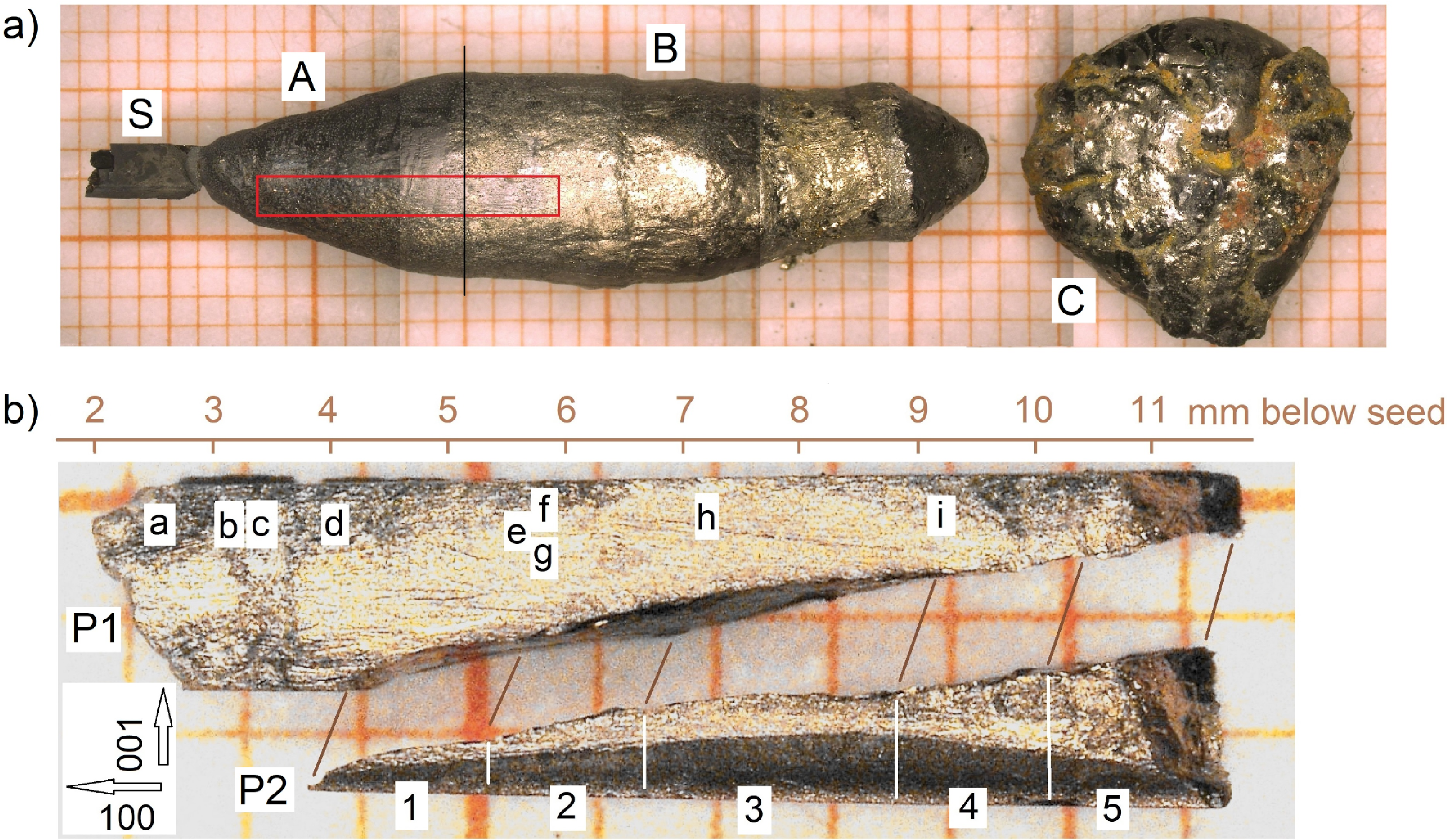}
	\caption[]{(a) Sample $\#3$ grown by the Czochralski method using a single crystalline seed S. 
The upper part A of the sample is free from macroscopic inclusions while the lower part B contains 
flux inclusions which are visible in the SEM picture. The residual flux C is not stable in air 
and decomposes after a few hours in air. (b) Single crystalline piece which was extracted from 
the sample (red box) shown in (a). The extracted crystal was divided into two different samples P1 and P2. 
Sample P1 was used for resistivity measurements at positions a-i and was characterized by single crystal analysis, EDX and WDX. Sample P2 was cut into  pieces 1-5 whose magnetization and heat capacity was measured.}
\label{CzochralskiProbe}
\end{figure}

\begin{figure}
\centering
\includegraphics[width=0.55\textwidth]{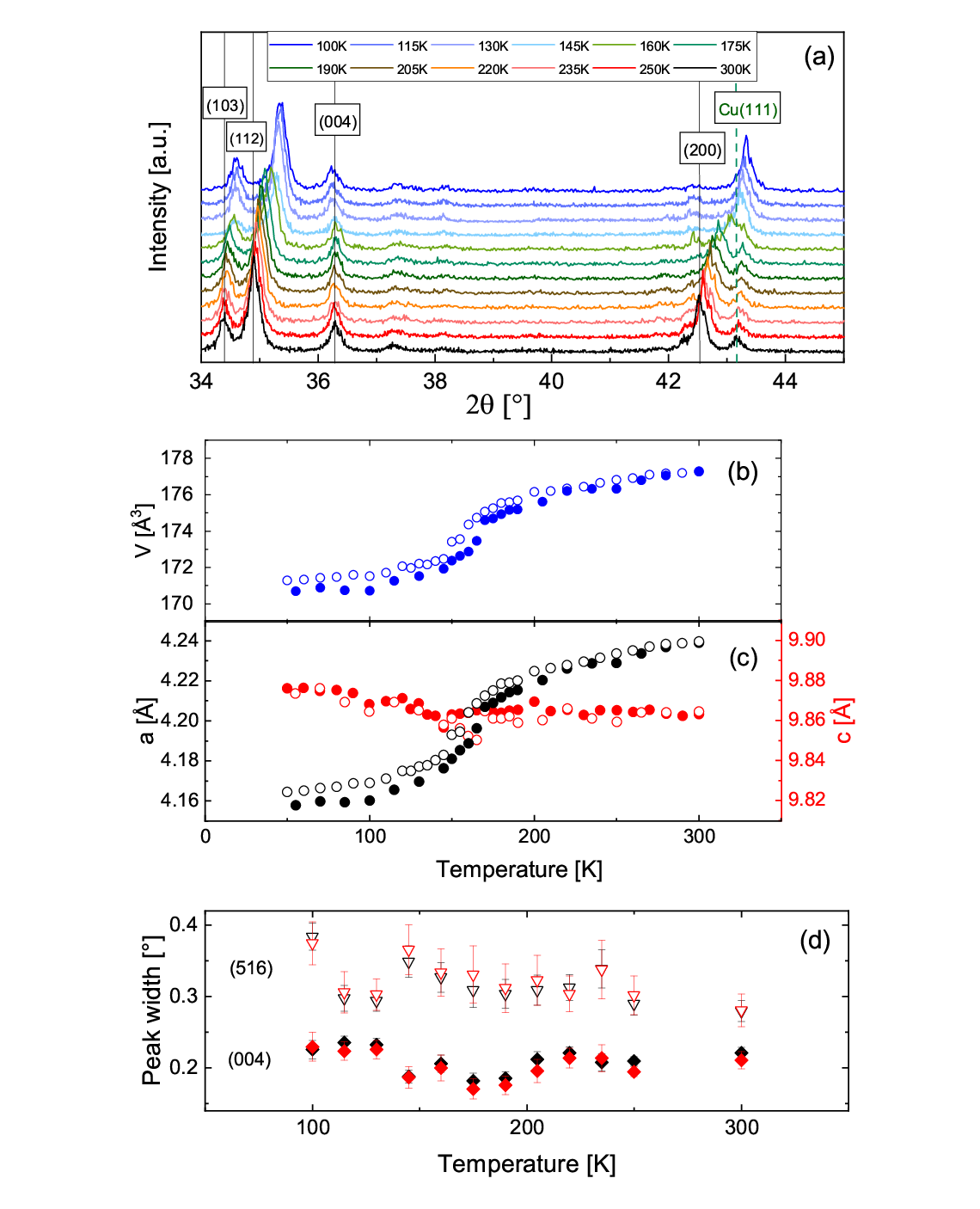}
	\caption[]{(a) Low-temperature powder X-ray diffraction between $100\,\rm K$ und $300\,\rm K$ 
performed on a Czochralski grown sample $\#2$. The green dashed line marks the position of the $(1\,1\,1)$ 
reflex of Cu from the sample holder which serves as standard. 
Change of the volume of the unit cell (b) and the lattice parameters (c) with temperature for Czochralski grown samples $\#1$ (closed symbols) and $\#2$ (open symbols). (d) FWHM of the $(5\,1\,6)$ and the $(0\,0\,4)$ reflex fitting a Gaussian (black symbols) or Lorentzian line shape (red symbols) for sample $\#1$.}
	\label{TTPXRD}
\end{figure}

\subsection{Crystal Growth}
The DTA experiments, Fig.~\ref{DTA_EC05}, showed that the compound solidifies in flux already at about 1200$^\circ$C at the initial weight stoichiometry of 
$\rm Eu :\rm Pd :\rm Si = 1.4:2:2$. This means that the growth temperatures can be 
chosen significantly lower than previously assumed and used for 
Bridgman growth \cite{Onuki2017} if a pre-reacting step is carried out. 
Both types of samples prepared by DTA and by Bridgman method showed a large amount of included secondary phases. 
Furthermore, these samples displayed a very broad transition in the specific heat above $100\,\rm K$. 
The valence transition temperatures which were determined by heat capacity 
and magnetization measurements exhibited strong sample dependencies.
During reproduction experiments using the Bridgman method, it was noticed that 
the grown crystals were on the one hand small (platelets, $\approx 1\,\rm mm \times 1.5\,\rm mm$), and on the other hand also contained 
many Eu-rich flux inclusions, which motivated us to develop the growth of this 
compound using the Czochralski method. 
The Czochralski growth of EuPd$_2$Si$_2$ is highly challenging: 
(i) Eu has a high vapor pressure at high temperatures, 
(ii) the melt attacks all tested crucible materials 
(tantalum, Al$_2$O$_3$, graphite, glassy carbon) at high temperatures, and 
(iii) the ideal starting stoichiometry for obtaining inclusion-free samples is unknown so far. 
We performed 10 crystal growth experiments each of which consisted of the pre-reaction steps 
followed by Czochralski growth. In all experiments, rod-shaped seeds were used with 
the long edges cut in a crystallographic direction perpendicular to the $c$-direction. 
Using these seeds, all samples were pulled out of the melt in a direction perpendicular to the $c$ direction. 
A typical growth result at the end of the optimization process is shown in Fig.~\ref{CzochralskiProbe}a).
During the optimization of the growth process, it was found that the evaporation 
of Eu from the melt can be slowed down sufficiently by an Ar overpressure of 20 bar in the growth chamber. 
 Due to the high reactivity of the melt, the growth was carried out from the levitating melt. 
We observed that the higher the Eu content, the larger the proportion of the target phase that can be drawn from the melt. 
The optimization process revealed that up to an initial stoichiometry of 
Eu$\,:\,$Pd$\,:\,$Si $= 1.45\,:\,2\,:\,2$, the EuPd$_2$Si$_2$ phase is the first to crystallize. 
Higher Eu contents lead to unstable conditions during the seeding phase which is possibly 
caused by the formation of a different phase at the beginning of the growth process. 
\begin{figure}
\centering
\includegraphics[width=0.5\textwidth]{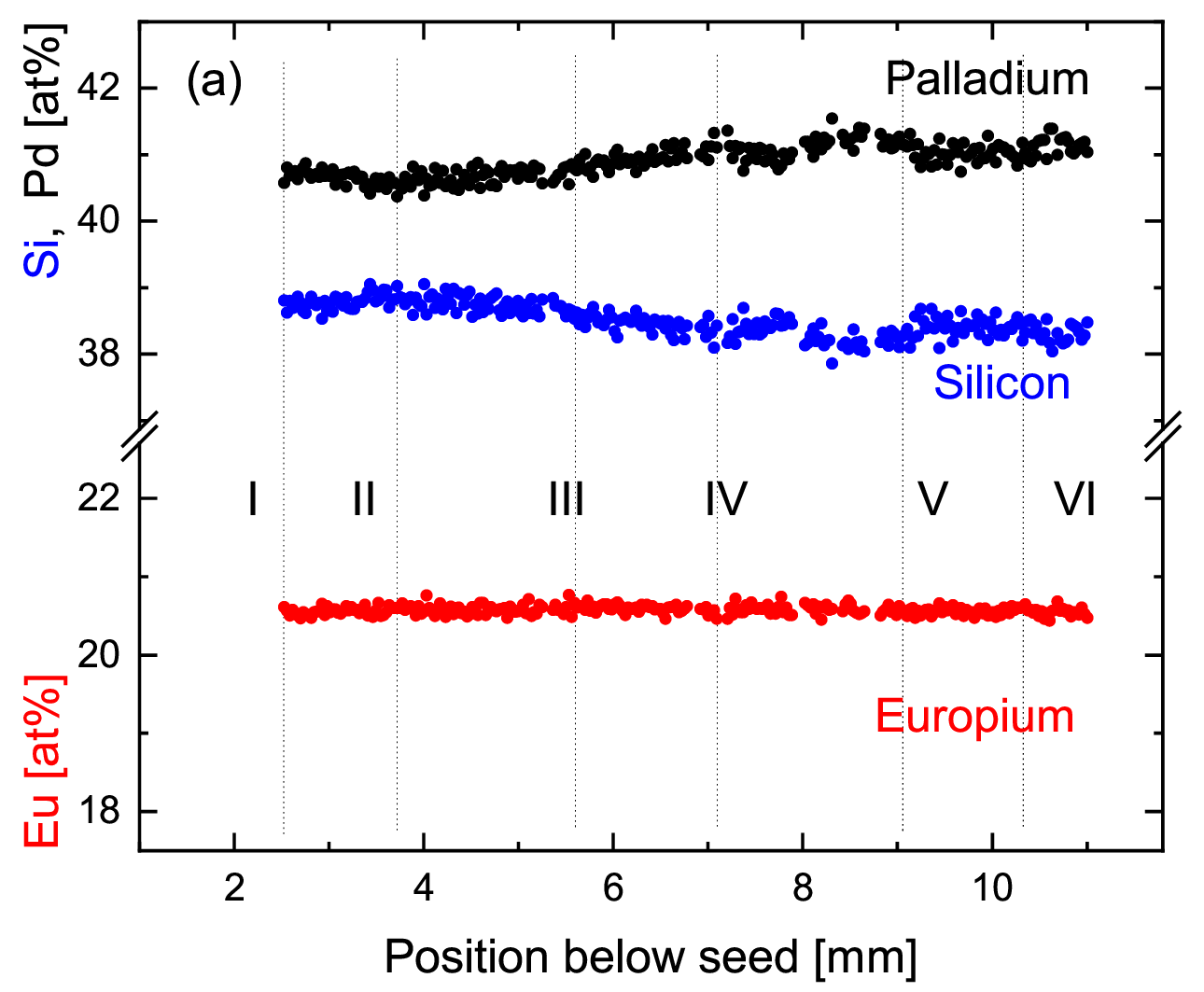}
\includegraphics[width=0.5\textwidth]{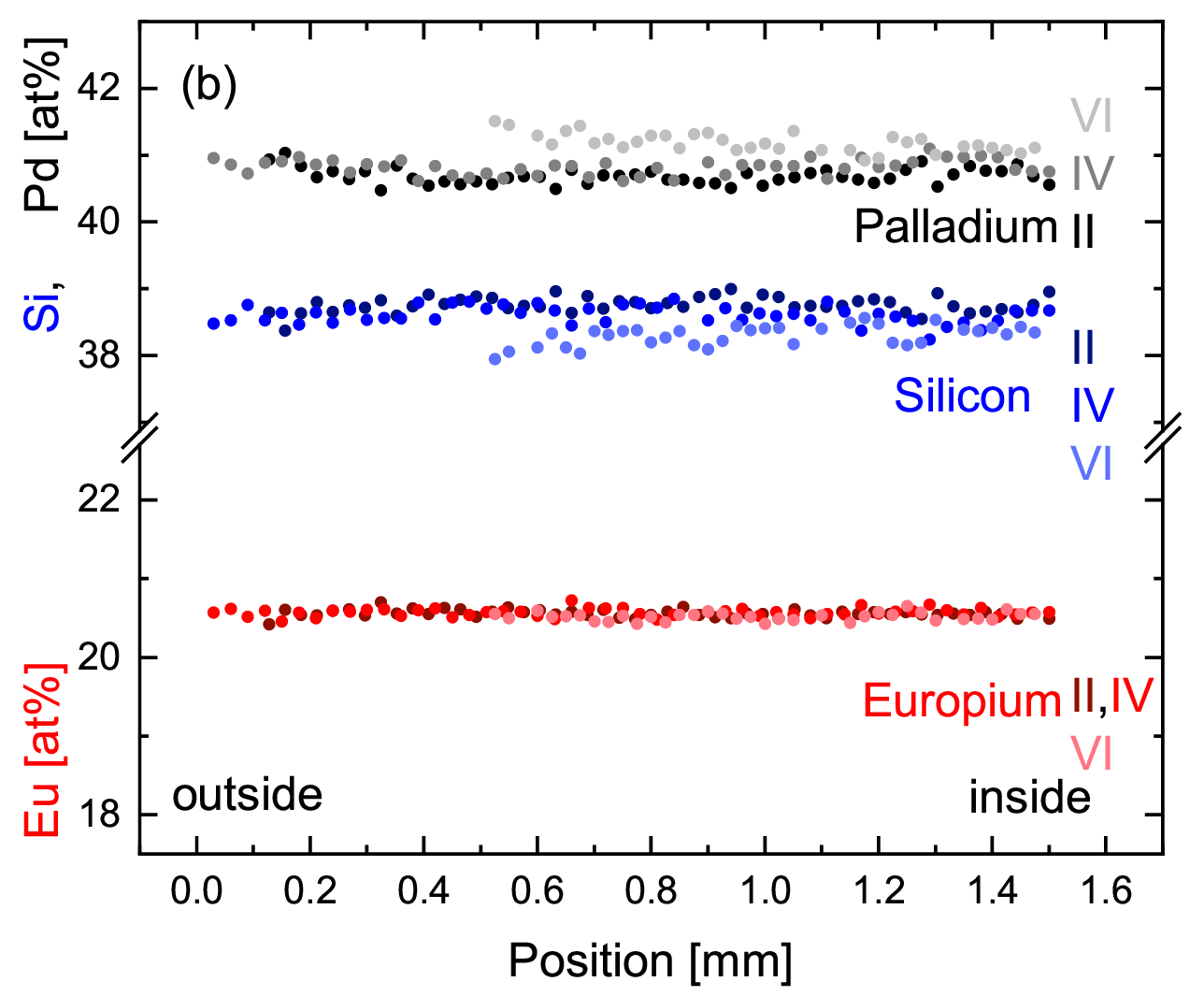}
\includegraphics[width=0.5\textwidth]{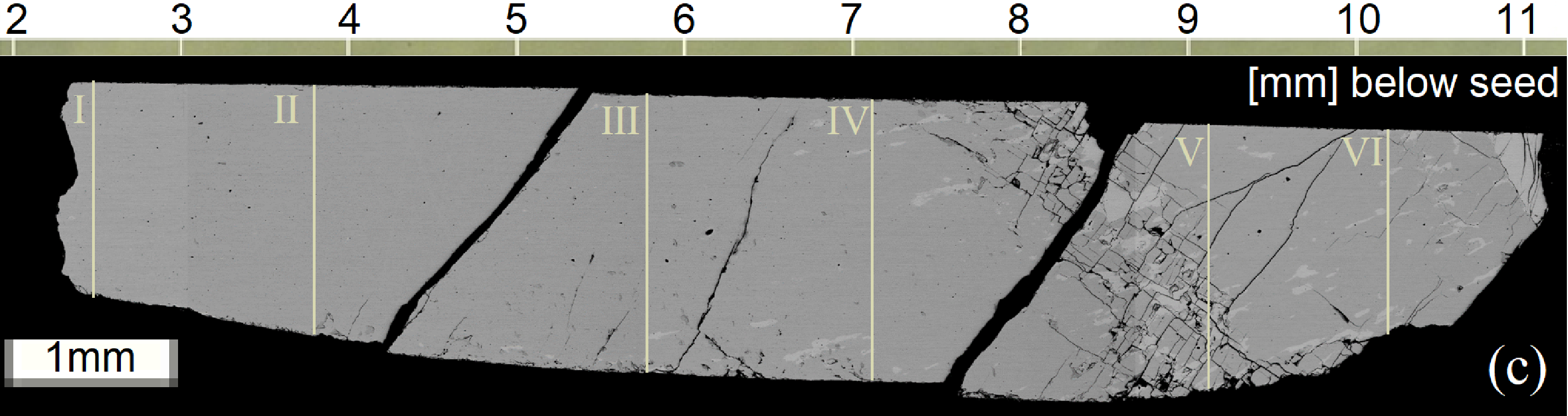}
	\caption[]{WDX was measured on the polished surface of (a)  a longitudinal cut through the sample P1, 
shown in Fig.~\ref{CzochralskiProbe}(b) and Fig.~\ref{WDX}(c). In the upper part of the sample (A) we observed a constant Eu content 
but a slight change of the Pd-Si ratio in the first $\approx 8.5\,\rm mm$. In the lower part (B) the sample 
contains macroscopic flux inclusions and the measured stoichiometry is not reliable. In (b) the analysis along the radial lines II, IV and VI is shown (positions marked in Fig.~\ref{WDX}(c)). (c) Electron microscope image of the polished sample which was analysed with WDX.}
\label{WDX}
\end{figure}

\subsection{Structural and Chemical Characterization}
PXRD confirmed the ThCr$_2$Si$_2$ structure of the samples with lattice parameters at $300\,\rm K$ of $a=4.2396(5)$\,\AA\, and $c=9.8626(4)$\,\AA\, which is in good agreement with literature \cite{Adams1991}. 
Low-temperature PXRD was performed on a Czochralski grown sample $\#2$ 
between 
$100\,\rm K$ and $300\,\rm K$, Fig.~\ref{TTPXRD}(a). For this sample, a large shift of the $(1\,0\,3)$, the $(1\,1\,2)$ 
and the $(2\,0\,0)$ reflections was observed between $120\,\rm K$ and $170\,\rm K$. 
The $(0\,0\,4)$ reflection exhibits only a small change of its position. Fig.~\ref{TTPXRD}(b,c) shows the temperature dependent volume and the lattice parameters of samples from two different Czochralski growth experiments $\#1$ (closed symbols) and $\#2$ (open symbols). 
The lattice parameter $a$ changes about $2\%$ while the length of the $c$ parameter remains nearly unchanged 
in the studied temperature range. The full width at half maximum (FWHM) of two reflections was analyzed 
by fitting their shapes with Gauss and Lorentz line shape. Our analysis of the $(5\,1\,6)$ and the $(0\,0\,4)$ reflections, shown in Fig.~\ref{TTPXRD}(d), yields no change of the FWHM upon cooling. 
From this we can conclude that no peak broadening of these reflections appears. 
Low-temperature PXRD was additionally performed on a sample grown by the Bridgman method (see supplemental information S1). In this case, we found 
a broadening of the $(1\, 1\, 2)$ peak between $140\,\rm K$ and $130\,\rm K$. Such a peak broadening was already observed 
in \cite{Jhans1987} and assigned to a distribution of $a$ lattice parameters in the sample. 
Since the physical properties reported in literature so far show a sample dependence, we performed 
the single crystal analysis, the WDX and the EDX analysis very carefully to find deviations from 
the ideal 122 stoichiometry. Prior to EDX analysis, we examined the sample by 
scanning electron microscopy (SEM) for visible flux inclusions. We found that the sample was 
free of inclusions up to a length of $\approx 8.5\,\rm mm$ below the seed.
The EDX analysis was performed on a longitudinal section for which a long rod-shaped single crystal (P1), Fig.~\ref{CzochralskiProbe}b,
was cut out of the grown sample $\#3$. This sample was finely polished for the analysis. 
The analysis of this longitudinal section
showed that the Eu content remained constant over the entire length of the sample studied. 
Furthermore, we found a small shift in the stoichiometry for Pd and Si within the first 
$8.5\,\rm mm$ of the sample. While the Pd content slightly increases, the Si content decreases at the same time.  
At a larger distance than $8.5\,\rm mm$ from the seed, no reliable statements can be made about 
the stoichiometry of the sample, since in this area the content of flux inclusions increases as it can be seen in Fig.~\ref{WDX}c. The analysis of the side phase stoichiometry can be found in the supplement S2.
The first hint for a slight change of the Pd-Si ratio in the grown samples came from the EDX analysis. 
Subsequently, to get a quantitative statement, we performed a WDX analysis on the same sample and 
could verify this observation for the longitudinal analysis, Fig.~\ref{WDX}a). At a distance of $3\,\rm mm$ below the seed, we determined a ratio of Eu$\,:\,$Pd$\,:\,$Si $= 20.5\,:\,40.5\,:\,39$, while a ratio of Eu$\,:\,$Pd$\,:\,$Si $= 20.5\,:\,41\,:\,38.5$ at a distance of $11\,\rm mm$ was found.
We additionally investigated radial lines of the sample, Fig.~\ref{WDX}b), and found the same trend.
Concerning the reproducibility, the occurring error is smaller than the size of the symbols used in Fig.~\ref{WDX}a,b.
To get a better insight in how this slight change of the composition might influence the bond lengths, we performed a single crystal analysis on small single crystals extracted from the same sample $\#3$.
The single crystal analysis yields that while for all samples the $4e$ Wyckoff position is completely occupied with Si, a certain amount of up to 3 \% Si is found on the Pd site (Wyckoff position $4d$). With decreasing Si content on the Pd site, the (Pd,Si)-Si bond length slightly decreases while the Eu-Si and the Si-Si distances along the crystallographic {\it c} direction are slightly increasing. Moreover, the $z$ parameter of the Si site (Wyckoff position $4e$) is directly correlated to the changes in the amount of Si on the Pd site (see Table \ref{tab:table1}). We like to mention that the observed differences in the structure are at the limit of what can be resolved with XRD and can be reliably given only in the context of measurements on a whole series of samples, as it was done here. \\
In our DTA, Bridgman and Czochralski growth experiments, we found different compositions 
of the included or remaining flux according to an EDX analysis. In DTA experiments, a silver-looking phase  
Eu$\,:\,$Pd$\,:\,$Si $= 26.4\,:\,28.1\,:\,45.5$ was found besides the main phase. 
In Bridgman growth experiments, the flux consisted of Eu,  
Eu$\,:\,$Pd$\,:\,$Si $= 1\,:\,1\,:\,1$ and Eu$\,:\,$Pd$\,:\,$Si $= 31.2\,:\,41.2\,:\,27.5$. 
In the Czochralski experiments we always found that the remaining material decomposed in air and the determination of its composition was impossible.

\begin{figure}
\centering
\includegraphics[width=0.5\textwidth]{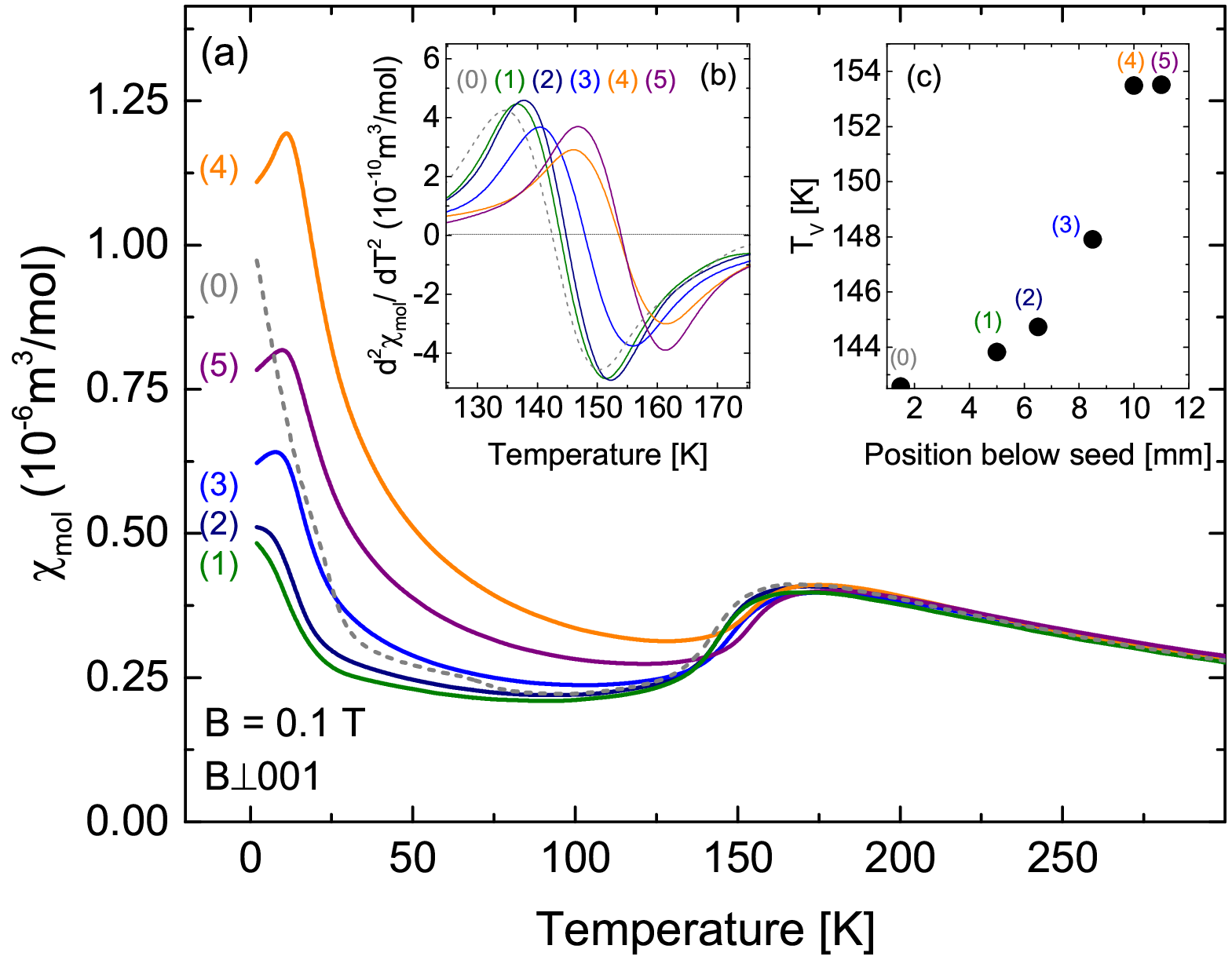}
	\caption[]{(a) The temperature dependence of the magnetic susceptibility was measured 
on different samples P2, (1)-(5) and (0). (b) The valence transition temperature $T_v$ was determined 
from the second derivative of the susceptibility. (c) $T_v$ depends on the extraction position 
of a sample from the grown crystal, see Fig.~\ref{CzochralskiProbe}(b).}
\label{Magnetisierung}
\end{figure}

\subsection{Magnetic susceptibility}

The magnetic susceptibility, Fig.~\ref{Magnetisierung}, measured with $B=0.1\,\rm T$, $B\perp c$, 
shows a strong change of the slope in the valence transition region. The valence transition temperature 
$T_v$ of a sample can be determined directly from the inflection point of the susceptibility curve. 
It is known from previous work that minor phases and impurities leave strong traces in the magnetization. 
Magnetization measurements are thus a sensitive tool to detect the presence of magnetic impurity phases. 
The magnetization of EuPd$_2$Si$_2$ was measured on 5 pieces, (1)-(5), whose exact extraction position 
from the sample is known (P2), Fig.~\ref{CzochralskiProbe}(b). From these measurements, we can conclude two different trends. i) The smaller the distance from the seed of a piece in the sample, the lower $T_v$. 
ii) The amount of impurity contaminations, seen at low temperatures, increases with larger distance from the seed. If the content of side phases is very high, 
then the side phase signal influences the susceptibility so strongly that the inflection point in 
the susceptibility at $T_v$ shifts or can no longer be determined.
The grey curve (0) in Fig.~\ref{Magnetisierung} exemplary shows the susceptibility of a 
sample extracted at a distance of about $0.5\,\rm mm$ below the seed. The data show that 
the valence transition temperature is lowest for this sample. Furthermore we observed for 
these early grown parts of the sample a change of the characteristics of the impurity contribution 
at low temperatures compared to pieces (1)-(5). Samples (1) and (2) show a small increase of the susceptibility below $20\,\rm K$ which might be assigned to a paramagnetic defect contribution \cite{Kemly1985}. Samples (3)-(5) contain a side phase with 
a magnetic transition at about $10\,\rm K$ which might correspond to the ordering of 
Eu$_2$PdSi$_3$ \cite{Mallik1998}. Contributions to the susceptibility below $100\,\rm K$ were already assigned to inclusions of magnetic side phases in this material in the past \cite{Holland1987}.

\begin{figure}
\centering
	\includegraphics[width=0.5\textwidth]{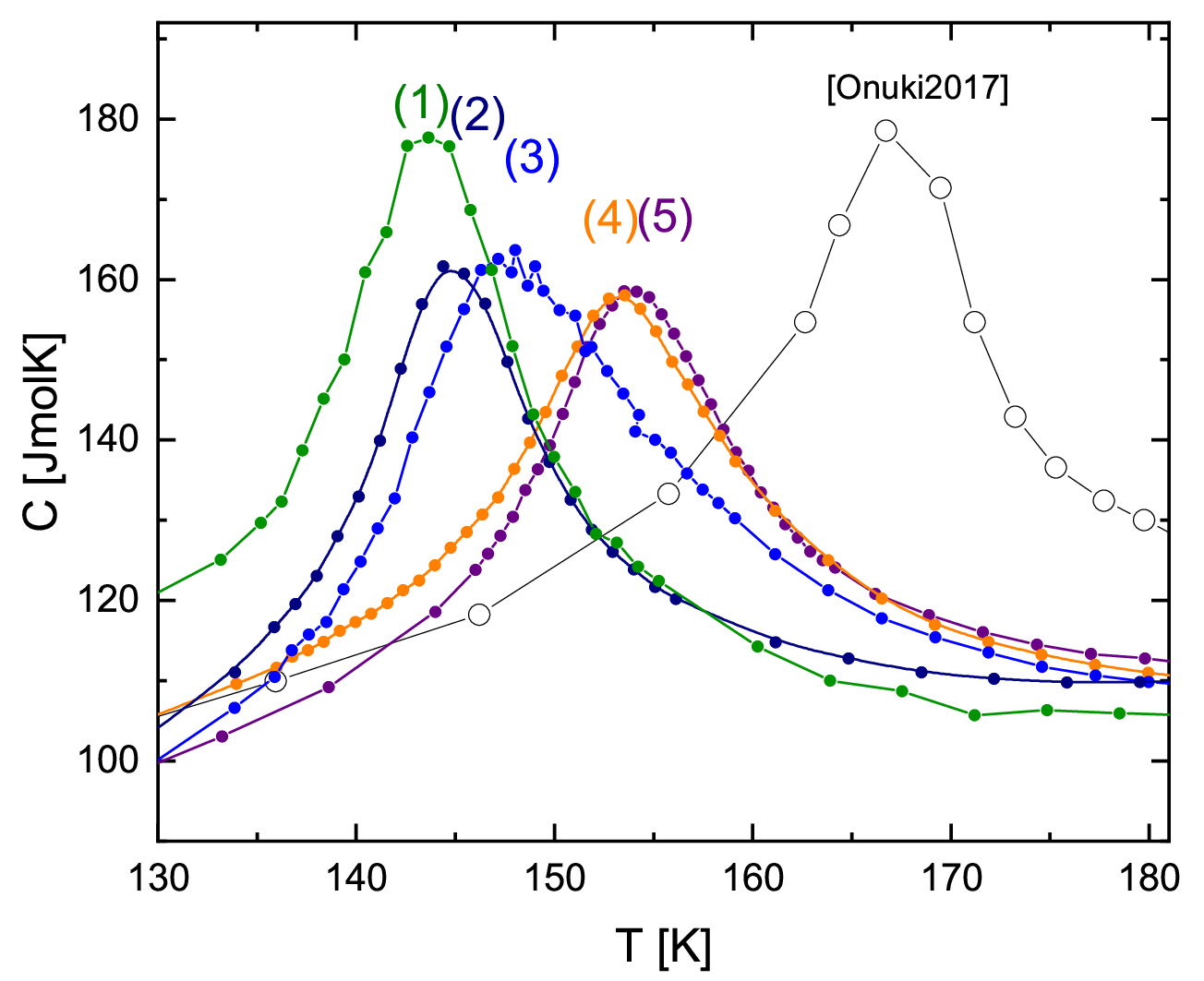}
	\caption[]{Temperature dependence of the heat capacity measured on samples P2, (1)-(5) 
(see Fig.~\ref{CzochralskiProbe}(b)) in comparison with data on single crystals from Ref.\cite{Onuki2017}.}
	\label{heatcapacity}
\end{figure}

\subsection{Heat capacity}
The heat capacity of EuPd$_2$Si$_2$ shows a peak at the valence transition. It is known from the 
literature that $T_v$ determined from heat capacity data is sample dependent: A polycrystalline sample showed 
$T_v=142\,\rm K$ \cite{Wada2001} while on a single crystalline sample $T_v=167\rm\, K$ 
was determined \cite{Onuki2017}.
To systematically investigate this sample dependence and also to look for latent heat as 
an indication of a 1$^{\rm st}$ order phase transition, we performed 
heat capacity measurements on the same samples (1)-(5) of piece P2, Fig.~\ref{CzochralskiProbe}(b), 
where the magnetization was measured. It was found that the position of the inflection point 
in the magnetization coincides with the position of the peak in the specific heat, Fig.~\ref{heatcapacity}.
Thus, in the heat capacity one also observes a dependence of the valence transition 
temperature on the position of extraction from the grown sample ranging from $T_v=142-154\,\rm K$. It has to be mentioned that in other Czochralski grown samples an even lower $T_v$ was detected.
None of the samples investigated so far showed signs of a 1$^{\rm st}$ order phase transition.

\begin{figure}
\centering
\includegraphics[width=0.5\textwidth]{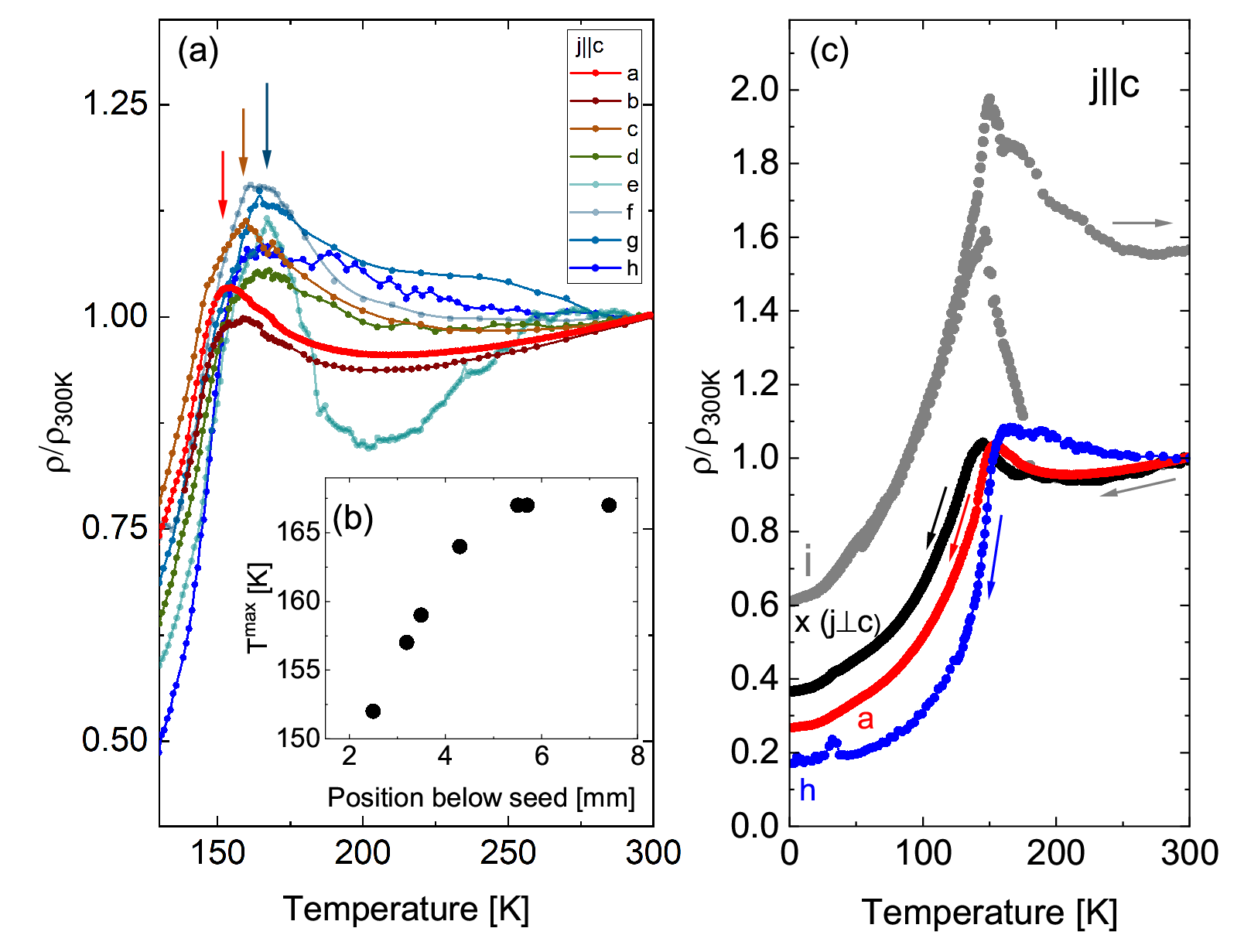}
	\caption[]{(a) The temperature dependence of the electrical resistivity was measured on different positions a-i 
(see Fig.~\ref{CzochralskiProbe}(b)) on one large single crystal P1 with current parallel to the $c$-direction. 
The arrows mark the position of the maximum which shifts for different contact positions. 
	(b) The temperature $T_{max}$ was determined from the maximum in the resistivity. (c) Electrical resistivity measured at contact positions a, h, i which shows a change of the shape of the curve. Curve x (black) shows the data obtained for $j\perp c$.}
\label{resistivity}
\end{figure}

\subsection{Electrical resistivity}

The temperature dependence of the electrical resistivity, $\rho(T)$, Fig.~\ref{resistivity},
was systematically measured on a long rod-shaped sample at different positions (a-i, Fig.~\ref{CzochralskiProbe}(b)) 
along the sample P1. The different positions of the contacts correspond to different distances from the seed. 
We observed that the measurements with current flow perpendicular to the $c$-direction succeeded only in 
a few cases. This was possibly related to a damage to the sample due to the strong change of the $a$ lattice parameter. 
In many cases it was not possible to obtain reliable data.
For position i which is already in the region with macroscopic flux inclusion we observe a different shape 
of the curve and a large hysteresis due to the possible damaging of the sample during the temperature change.
However, the measurement of the resistivity 
with $j\parallel c$ was possible in most cases. 
We observed  a shift in the maximum for different contact positions a-h on sample P1, 
Fig.~\ref{resistivity}(a). The temperature at which the maximum occurs, Fig.~\ref{resistivity}(b), depends 
on the contact position.
The resistivity ratio RR$_{1.8\,\rm K}=\rho(300\,\rm K)/\rho(1.8\,\rm K)$ varies between RR$_{1.8\,\rm K}=3.7$ (position (a)) and RR$_{1.8\,\rm K}=5.9$ (position (h)). 
For $j\parallel c$, we observed a small increase in the resistivity at $T_v$ followed by 
a sharp drop towards low T during cooling. The curve shape between the two current directions differs only slightly. For $j\perp c$, black curve, x, in Fig.~\ref{resistivity}(c), the resistivity curve shows a similar characteristics at $T_v$ but a lower residual resistivity ratio of RR$_{1.8\,\rm K}\approx 2.5$.

\section{Discussion}
The growth of Eu-based systems as single crystals is challenging mainly due to the incongruent solidification of the materials and due to the strong tendency of Eu to oxidize. In this work, we developed a process for growing crystals of EuPd$_2$Si$_2$ by the Czochralski method. We found that with an initial stoichiometry of 
Eu$\,:\,$Pd$\,:\,$Si $= 1.45\,:\,2\,:\,2$, the EuPd$_2$Si$_2$ phase is the first to crystallize from the melt. 
We performed the structural and chemical characterization to compare the properties of our samples with those reported in literature for powder samples and Bridgman grown single crystals. 
Low temperature PXRD experiments, performed between $300\,\rm K$ and $10\,\rm K$ 
on Czochralski grown samples, showed that the $a$ parameter undergoes a large change of 
about 2\% upon cooling, while the $c$ parameter remains almost constant. 
The valence transition manifests itself in a strong shift of the reflections, which can be observed 
e.g. at the $(2\,0\,0)$ reflection. In contrast, for a Bridgman grown sample we found a peak broadening for the $(2\,0\,0)$ reflection similar as it was reported in \cite{Jhans1987}. 
For the purer Czochralski grown samples, no peak broadening is observable which shows that this is not an intrinsic property of this material.
The chemical characterization by WDX yielded that during the Czochralski growth process the compound solidifies in a homogeneity range Eu(Pd$_{1-m}$Si$_m$)$_2$Si$_2$. The analysis confirmed small changes in the Pd:Si ratio in the order of $1\,\rm at.\%$ along the growth direction of the crystal. Although the systematic error in the absolute values might be larger than that change the relative trend is robust among different measurements. To get more microscopic insight into the structural variations, in parallel we conducted a single crystal analysis on samples which were extracted from the same grown ingot. 
The careful analysis of different single crystals showed that the Pd site is partially occupied by Si which is accompanied by a small change of the Si position and correspondingly in the bond distances.
The small change in the Pd-Si ratio accompanied by the change in the bond lengths, has a large impact on the temperature $T_v$ at which the valence transition occurs.
In the electrical resistivity, measured with current parallel to the $c$ direction on the same crystal used for the WDX analysis, 
we found maxima between $152\,\rm K \leq T_{\rm max}\leq 167\,\rm K$. The neighbouring part of that crystal was used to determine the valence transition from thermodynamic quantities.  
We determined $142\,\rm K \leq T_v\leq 154\,\rm K$ from the inflection point in the 
magnetic susceptibility and heat capacity measurements showed that the peak temperature corresponds to $T_v$ determined from the magnetic susceptibility. From this thorough analysis of the structural and physical properties, we conclude that with increasing distance to the seeding crystal, the stoichiometry of EuPd$_2$Si$_2$ moves towards the ideal 122 composition, accompanied by an increase of $T_v$ and an increase of the residual resistivity ratio. However, among our samples we could not observe crystals with an ideal 122 stoichiometry, as there remains some Si excess on the Pd site. In all cases the transition at $T_v$ rather is a crossover than a second order phase transition, and  
none of the samples showed signs of a $1^{\rm st}$ order transition. This is consistent with EuPd$_2$Si$_2$ being 
located at the high pressure side, very close to the critical endpoint of a line of 
first order phase transitions in a general $p-T$ diagram for Eu compounds \cite{Onuki2017}.

\section{Summary}
We showed that the intermediate valent, tetragonal EuPd$_2$Si$_2$ can be grown from a Eu-rich levitating melt using the Czochralski method with an Ar overpressure of $20\,\rm bar$. 
The combined chemical and structural analysis revealed that the material grows in a Pd-Si homogeneity range and that this small change of the order of $1\,\rm at.\%$ in the composition along the grown ingot of the material has a large impact on the temperature where the valence transition occurs. In this sample, we observe a shift in $T_v$ of about $15\,\rm K$. Samples with lower $T_v$ formed at the beginning of the growth while samples with higher $T_v$ formed later in the process. 
The here presented growth procedure using the Czochralski method will now enable us 
to provide doped samples and chemically tune this material to further explore the phase diagram and it shows that EuPd$_2$Si$_2$ is an excellent candidate for the study of possible critical elasticity.

\section*{Acknowledgements}
We thank K.-D. Luther for technical support and R. M\"oller for proofreading the manuscript. We acknowledge funding by the Deutsche Forschungsgemeinschaft (DFG, German Research Foundation) via the TRR 288 (422213477, projects A03 and B03).

\noindent {\bf Supporting Information:} 
{\bf S1} Low-temperature PXRD data of a Bridgman-grown sample. {\bf S2} Analysis of the included side phase in a Czochralski-grown sample.

\bibliography{EuPd2Si2_Bibtex}
\clearpage
\section{Supporting information}
\subsection{S1 Low-temperature powder X-ray diffraction Bridgman-grown sample}          
 \begin{figure}
\centering
\includegraphics[width=0.5\textwidth]{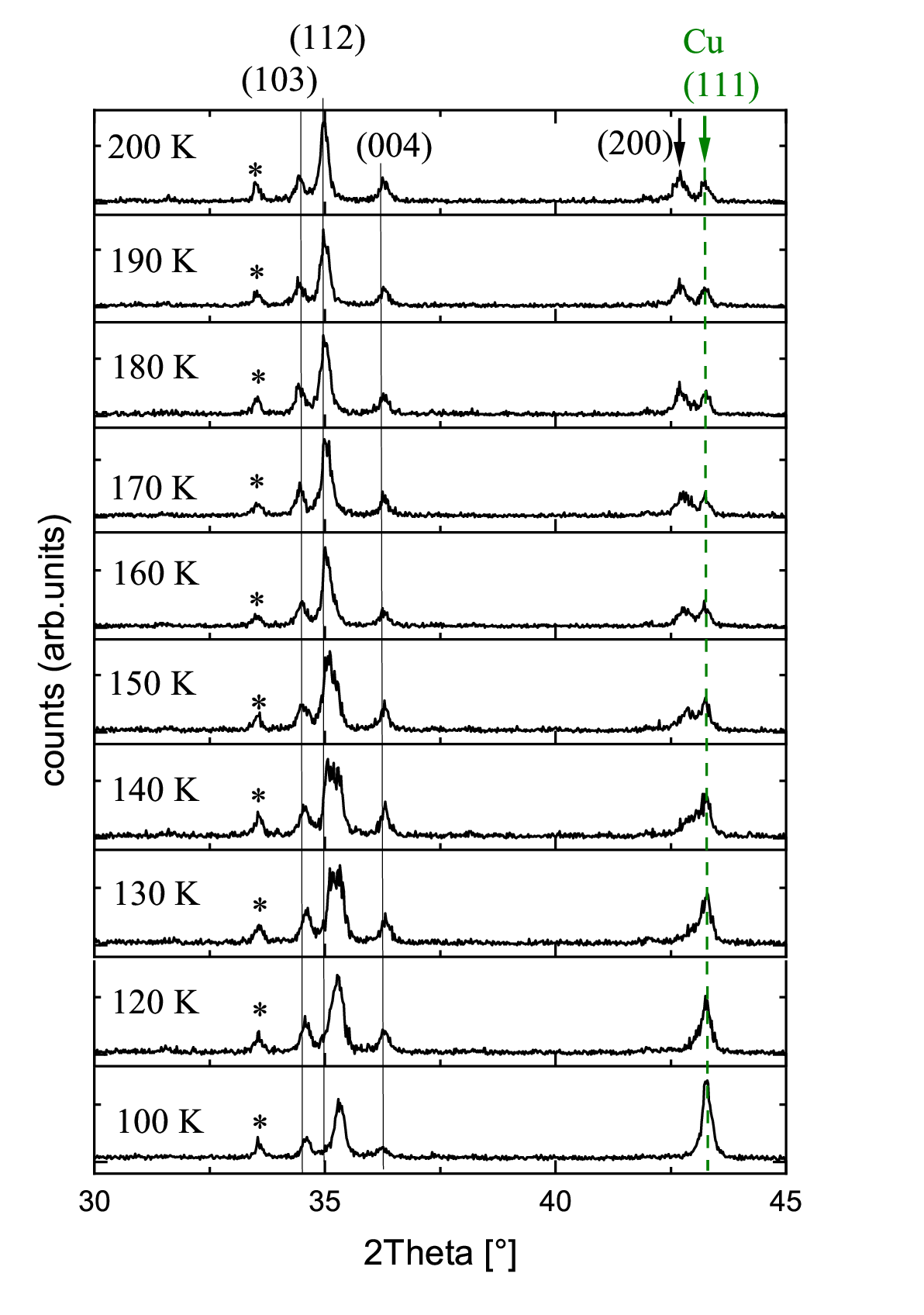}
	\caption[]{Low-temperature powder X-ray diffraction between $100\,\rm K$ und $200\,\rm K$ 
performed on a Bridgman-grown sample (EC06). The green dashed line marks the position of the $(1\,1\,1)$ 
reflex of Cu from the sample holder which serves as standard. The asterisk marks a reflex which belongs to an unidentified impurity phase.}
	\label{TTPXRD_S1}
\end{figure}         
 
 \newpage
 \subsection{S2 Analysis of the included side phase in a Czochralski-grown sample}
 
The side phase inclusions in our sample $\#3$ were analysed using WDX. We found a stoichiometry of Eu$_{13}$Pd$_{18}$Si$_{19}$ in average which does not correspond to a known phase in the Eu-Pd-Si system. 
 \begin{figure}
\centering
\includegraphics[width=0.5\textwidth]{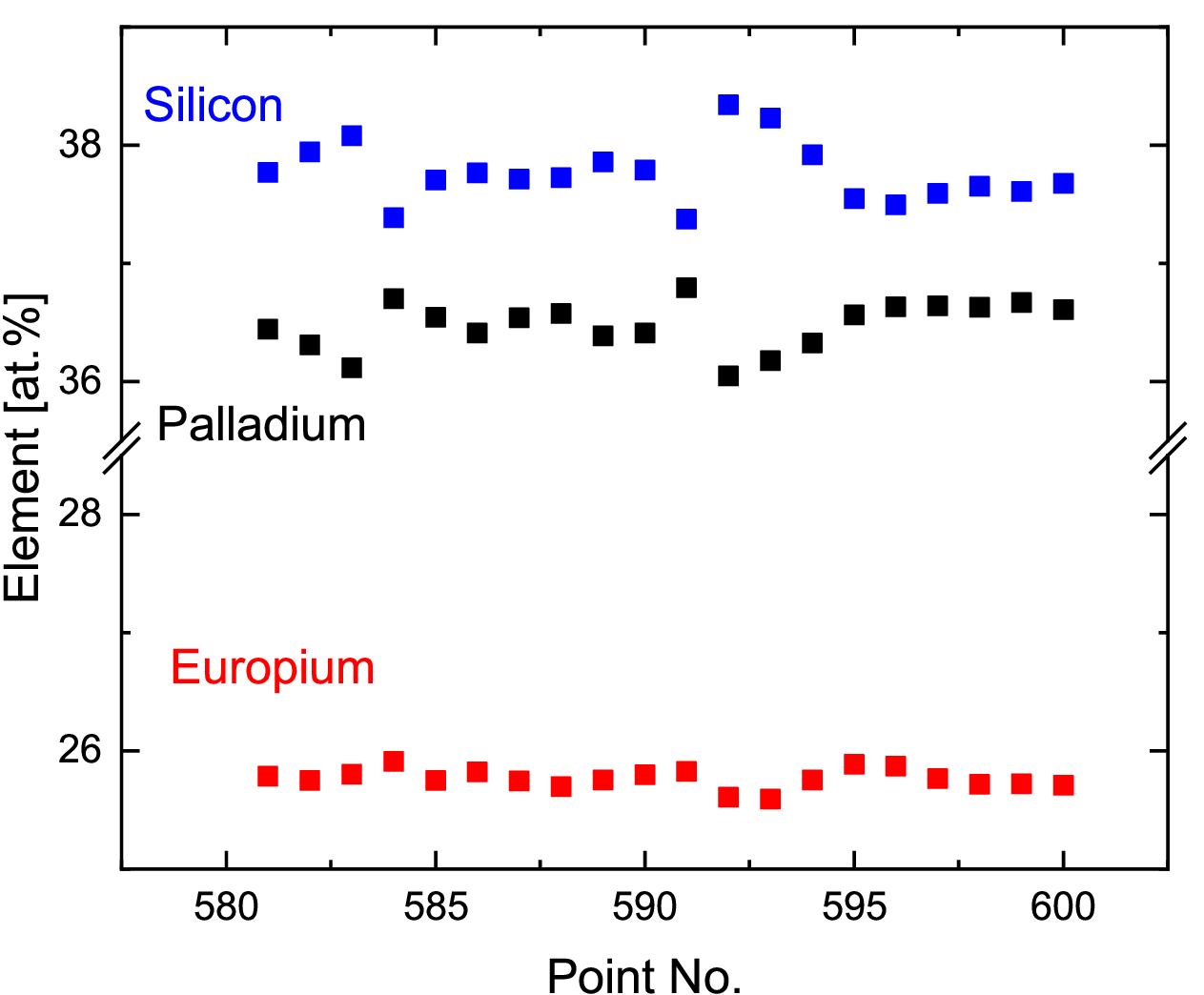}
	\caption[]{WDX analysis of the included side phase in the Czochralski-grown sample $\#3$.}
	\label{TTPXRD_S1}
\end{figure}  

\end{document}